\documentclass[10pt]{article}
\usepackage[utf8]{inputenc}
\usepackage{amsfonts,amsmath,amssymb}
\usepackage{graphicx}
\usepackage{bm}
\usepackage{braket}
\usepackage{hyperref}
\usepackage{array}
\usepackage{epstopdf}
\usepackage{authblk}

\newcolumntype{C}[1]{>{\centering\let\newline\\\arraybackslash\hspace{0pt}}m{#1}}

\newcommand{\D}{\mathrm{d}}
\newcommand{\Hb}{\overline{\mathrm{H}}}
\newcommand{\gb}{\overline{g}}
\newcommand{\calS}{\mathcal{S}}
\newcommand{\Hart}{\mathrm{E}_\mathrm{h}}
\newcommand{\Bohr}{\mathrm{a}_0}

\newcommand{\WKB}{{\scriptscriptstyle \mathrm{WKB}}}

\begin{document}

\title{Quantum Reflection of Antihydrogen in the GBAR Experiment}

\author[1]{G.~Dufour}
\author[1]{R.~Guérout}
\author[1]{A.~Lambrecht}
\author[2]{V.V.~Nesvizhevsky}
\author[1]{S.~Reynaud}
\author[3]{A.Yu.~Voronin}

\affil[1]{Laboratoire Kastler-Brossel, CNRS, ENS, UPMC, Campus Jussieu, F-75252 Paris, France}
\affil[2]{Institut Max von Laue - Paul Langevin, 6 rue Jules Horowitz, F-38042, Grenoble, France }
\affil[3]{P.N. Lebedev Physical Institute, 53 Leninsky prospect, Ru-117924 Moscow, Russia }

\date{}

\maketitle

\begin{abstract}
In the GBAR experiment, cold antihydrogen atoms will be left to fall on an annihilation plate with the aim of measuring the gravitational acceleration of antimatter.
Here, we study the quantum reflection of these antiatoms due to the Casimir-Polder potential above the plate. 
We give realistic estimates of the potential and quantum reflection amplitudes, taking into account the specificities of antihydrogen and the
optical properties of the plate. We find that quantum reflection is enhanced for weaker potentials, for example above thin slabs, graphene and nanoporous media.

\vspace{0.3cm}


\end{abstract}

\section{Introduction}

The GBAR collaboration (Gravitational Behavior of Antihydrogen at Rest) aims to measure the gravitational behavior of antimatter by timing the free fall of antihydrogen ($\Hb$) atoms\cite{chardin2011,perez2012,debu2012}. The proposed method\cite{walz2004} is to trap $\Hb^+$ ions and cool them down to the lowest quantum state in a Paul trap. $\Hb$ is then produced by photo-detaching the excess positron. The photo-detachment pulse is the START signal for the free fall timing measurement, while the STOP signal is provided by the annihilation of $\Hb$ atoms on a detection plate placed at a height $h$ below the ion trap.
The acceleration of gravity $\gb$ of $\Hb$ in the Earth's gravity field is then deduced from the distribution of free fall times, assuming no other forces are acting on the neutral atom.

Casimir and Polder have shown\cite{casimir1946,casimir1948} that neutral atoms in the vicinity of a material medium experience an attractive force because the atomic induced dipole is coupled to induced dipoles in the material via electromagnetic vacuum fluctuations.
Although the effect of this force on the free fall time is well below experimental accuracy, the Casimir-Polder (CP) interaction is known to cause quantum reflection of atoms at low energies\cite{echenique1976,mody2001,friedrich2002,dufour2013}.

Classically forbidden reflection of a particle from an attractive potential is a well known effect in quantum mechanics; it occurs when the potential varies rapidly on the scale of the de Broglie wavelength of the particle\cite{berry1972,mody2001,friedrich2002}.
Several experiments have observed such reflection on the CP potential near liquid \textrm{He}\cite{nayak1983,berkhout1989,yu1993} and solid surfaces\cite{shimizu2001,druzhinina2003,pasquini2004}, as well as rough or micro-/nanostructured surfaces\cite{shimizu2002,pasquini2006,zhao2010}.

In this paper we use the scattering approach\cite{lambrecht2006,emig2007} to calculate the interaction of $\Hb$ with various types of surfaces.
We then calculate the associated quantum reflection, taking into account the fact that $\Hb$ annihilates when it comes in contact with matter\cite{voronin2005-a,voronin2005}. This means that in contrast with normal matter atoms, the short range atom-surface interactions do not enter the problem. 
Interpreting quantum reflection as a deviation from the semiclassical Wentzel-Kramers-Brillouin (WKB) approximation allows us to explain why larger reflection is obtained for weaker potentials. In order to take advantage of this increased reflection, we consider several materials that are weakly coupled to the electromagnetic field.

\section{Casimir-Polder potential}

In the scattering approach, the CP potential is expressed in terms of the electromagnetic reflection operators on each of the interacting objects\cite{messina2009,dufour2013}:
\begin{itemize}
\item reflection on the plane is described by the Fresnel reflection amplitudes, which depend on the relative dielectric function of the medium,
\item reflection on the atom depends on its dynamic polarizability $\alpha(\omega)$, which is supposed to be the same as that of (ground state) Hydrogen.
\end{itemize}
This formalism allows an easy inclusion of realistic optical response properties for the atom and material slab\cite{lambrecht2007}. Those used in this paper are detailed in Ref. \cite{dufour2013}.
Since quantum reflection occurs at distances smaller than $1\mu$m, which is the typical thermal wavelength, all calculations are performed for
zero temperature.

The typical wavelength $\lambda$ characterizing the optical response of the atom and plane defines the transition between two asymptotic behaviors of the CP potential:
\begin{equation}
\label{vlimits} V(z) \underset{z \ll \lambda}{\to} -\frac{C_3}{z^3} \quad,\quad V(z)\underset{z \gg \lambda}{\to} -\frac{C_4}{z^4}~.
\end{equation}
The short distance limit is well known as the van der Waals potential; whereas the large separation limit is referred to as the \emph{retarded} CP interaction since it takes into account the finiteness of the speed of light\cite{casimir1946,casimir1948}.

\begin{figure}[th]
\centering
 \includegraphics[width=6.5cm]{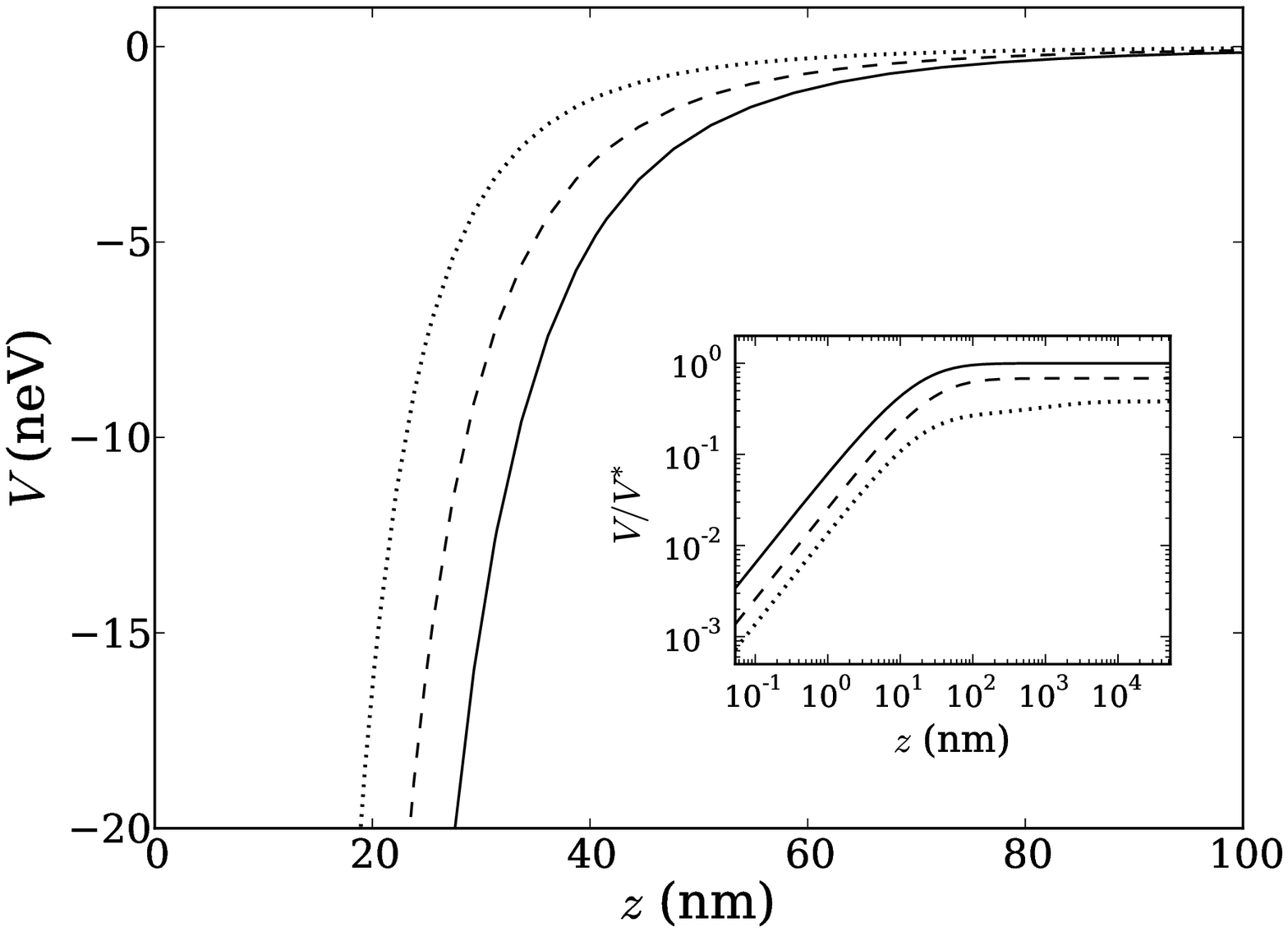}\hspace{-.5cm} \includegraphics[width=6.5cm]{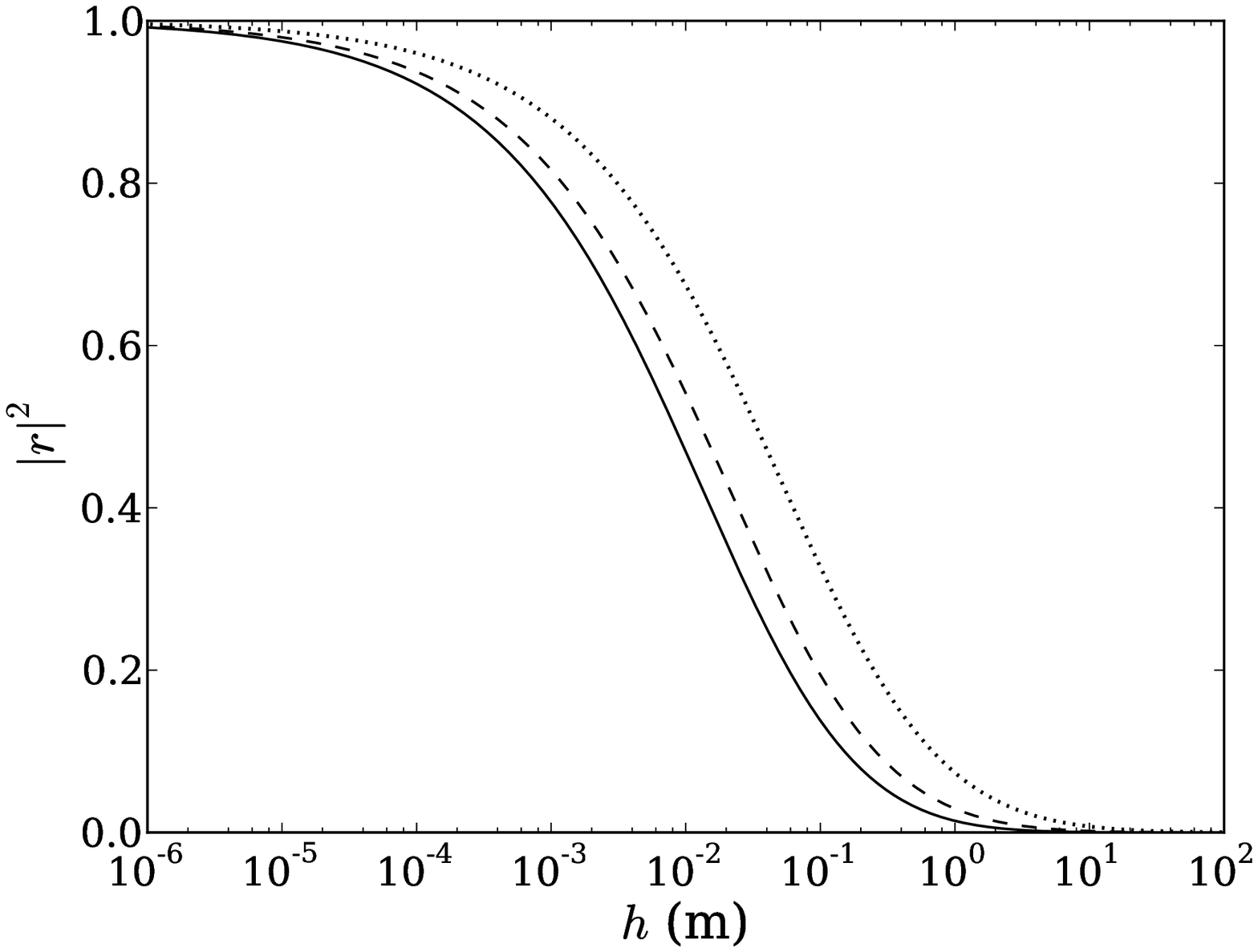}
\caption{\label{pot-ref}Left: CP potential for $\Hb$ in the vicinity of a
material bulk; from top to bottom, perfect conductor
(full line), silicon (dashed line), silica (dotted line); (inset: ratio $V/V^*$ to the retarded potential
$V^*$ for a perfectly conducting mirror, see text).\newline
Right: Quantum reflection probability $\left\vert
r\right\vert^2$ as a function of the free fall height $h$ for $\Hb$
atoms on bulk mirrors; from bottom to top, perfect conductor (full line),
silicon (dashed line), silica (dotted line).}
\end{figure}

The left plot in figure \ref{pot-ref} displays the exact CP potentials between an $\Hb$ atom and bulk mirrors made of a perfectly conducting metal, intrinsic silicon or amorphous silica.
The inset shows the ratios $V(z)/V^*(z)$ to the retarded CP limit calculated for a perfect conductor: $V^*(z)=-C_4^*/z^4$, with
$C_4^*= (3\hbar c/ 8 \pi)(\alpha(0)/4\pi\epsilon_0)$, $C_4^* = 1.57 \times 10^{-7}$~neV.nm$^4$=73.6 $\Hart$.$\Bohr^4$ (Hartree energy $\Hart=4.3597$~aJ, Bohr radius $\Bohr=52.917$~pm). 
These ratios tend to constant values $C_4/C_4^*\leq1$ at large distances and linear variations $C_3 z/C_4^*$ at small distances. 
The less reflective for the electromagnetic field a material is, the weaker the CP potential, from perfect conductor to silicon and silica mirrors.

The values of $C_3$ and $C_4$ obtained from the exact CP potential for a perfect conductor, silicon and silica bulks are given in table \ref{c3c4}.

\begin{table}[h]
\caption{Numerical values of the $C_3$ and $C_4$ coefficients for hydrogen or antihydrogen atoms above bulk mirrors.}
{\label{c3c4}\begin{tabular}{C{2.5cm} |C{1.5cm} | C{2.5cm}|C{1.5cm} | C{2.5cm}} \hline \hline
mirror & $C_3$ ($\Hart.\Bohr^ 3$) & $C_3$ ($10^6$ neV.nm$^3$) & $C_4$ ($\Hart.\Bohr^ 4$)& $C_4$ ($10^7$ neV.nm$^4$)\\ \hline
perfect conductor         & 0.25  & 1.01 & 73.6  & 1.57 \\ \hline
bulk silicon              & 0.10  & 0.41 & 50.3  & 1.07 \\ \hline
bulk silica               & 0.05  & 0.21 &  28.1 & 0.60 \\ \hline
 \hline
\end{tabular}
}
\end{table}

\section{Quantum reflection of $\Hb$ }

We can now write the Schrödinger equation for the atom's vertical motion in the potential calculated in the previous section:
\begin{equation}
\label{schrod} \psi^{\prime\prime}(z) +\frac{p^2(z)}{\hbar^2} \psi =0~, \qquad \qquad p(z)=\sqrt{2m\left( E-V(z) \right)}~.
\end{equation}
Primes denote derivation with respect to $z$, $p$ is the semiclassical momentum and $E>0$ is the kinetic energy of the atom before it reaches the potential. To make the connection with the free-fall problem, we will often use the free fall height $h$ as a measure of the energy $E=mgh=102.5$~neV/m $\times h$.

To underline the effect of quantum reflection, we write the wavefunction in the basis of WKB waves, which each propagate in a well defined direction:
\begin{equation}\label{wkbwaves}
\psi \left( z\right) = \frac{c_+(z)}{\sqrt{\left\vert p(z)\right\vert}} e^{i\phi\left( z\right)} +
\frac{c_-(z)}{\sqrt{\left\vert p(z) \right\vert}} e^{-i\phi\left(z\right)}~, 
\qquad  \phi \left( z\right) = \int_{z_0}^{z} \frac{p (z^\prime) \D z^\prime}\hbar~,
\end{equation}
where $\phi$ is the WKB phase ($z_0$ arbitrary).

Introducing this ansatz in \eqref{schrod} we obtain coupled first-order equations for the amplitudes $c_\pm(z)$, which describe the conversion of an incident wave into a reflected wave and vice-versa\cite{berry1972}:

\begin{equation}\label{cpcm}
c_\pm^{\prime}(z) = e^ {\mp 2i \phi \left( z\right) }\,\frac{p^\prime(z)}{2 p(z)} \,c_\mp(z) ~.
\end{equation}

Because $\Hb$ annihilates if it touches the wall, there can be no outgoing wave immediately above the material surface.
This enforces a full absorption boundary condition $c_+(z=0)=0$.
Analytical solutions of Eqs. \eqref{cpcm} obeying that condition are available near the origin\cite{voronin2005-a,voronin2005,dufour2013}. They can be used as boundary conditions for the numerical integration of \eqref{cpcm} to avoid problems arising from the divergence of the potential.
The ratio of the amplitudes $c_+(z)$ and $c_-(z)$ as $z$ goes to infinity is the quantum reflection amplitude $r$. 

The right plot of Fig. \ref{pot-ref} shows the reflection probability $\left\vert r\right\vert^2$ as a function of the energy for each of the potentials calculated in the previous section. Significant values are obtained for an energy $E=m g \times 30$ cm typical of GBAR: the reflection probability is 5\% on a perfect conductor, 9\% on bulk silicon and 18\% on bulk silica.

Surprisingly, comparison of the left and right plots of Fig. \ref{pot-ref} shows that the reflection probability is larger for weaker CP interactions\cite{judd2011,pasquini2006,mody2001}. We explain this behavior in the next section by looking more closely at the condition for efficient reflection.

\section{Badlands condition}

We have seen in the previous section that quantum reflection can be introduced naturally as a deviation from the semiclassical WKB approximation, resulting in an exchange between the otherwise decoupled WKB waves.

If the amplitudes $c_\pm$ are \emph{not} allowed to vary, the wavefunction $\psi_\WKB$ obeys a modified Schrödinger equation:
\begin{equation}\label{schrodwkb}
  \psi_\WKB^{\prime\prime}(z) + (1+Q(z))\frac{p^2(z)}{\hbar^2} \psi_\WKB(z) = 0~, \qquad Q(z) = \frac{\hbar^2\calS\phi}{2p^2}
\end{equation}
The difference between \eqref{schrodwkb} and \eqref{schrod} is the extra term $Q(z)$, proportional to the Schwarzian derivative of the WKB phase:
\begin{equation}
\calS\phi(z) \equiv
\frac{\phi^{\prime\prime\prime}(z)}{\phi^{\prime}(z)} - \frac32
\left( \frac{\phi^{\prime\prime}(z)}{\phi^{\prime}(z)} \right)^2
=\frac{p^{\prime\prime}(z)}{p(z)} - \frac32
\left( \frac{p^\prime(z)}{p(z)} \right)^2~.
\end{equation}

In regions where $|Q(z)|$ is much smaller than one, the WKB approximation is good and there is no reflection ($c_\pm$ remain constant). Conversely, in regions where $|Q(z)|$ takes values of order one we can expect significant quantum reflection\cite{berry1972,maitra1996,cote1997,friedrich2002}; these are the so-called \emph{badlands}. In the case of the CP potential, one can show that the function $|Q(z)|$ is peaked around the region where $|V(z)|\simeq E$ and vanishes both at infinity and near the surface. Indeed at infinity the potential vanishes and close to the surface the momentum of the particle is very large, so that in both limits the behavior of the atom is classical.

\begin{figure}[thb]
\centering
 \includegraphics[width=6.5cm]{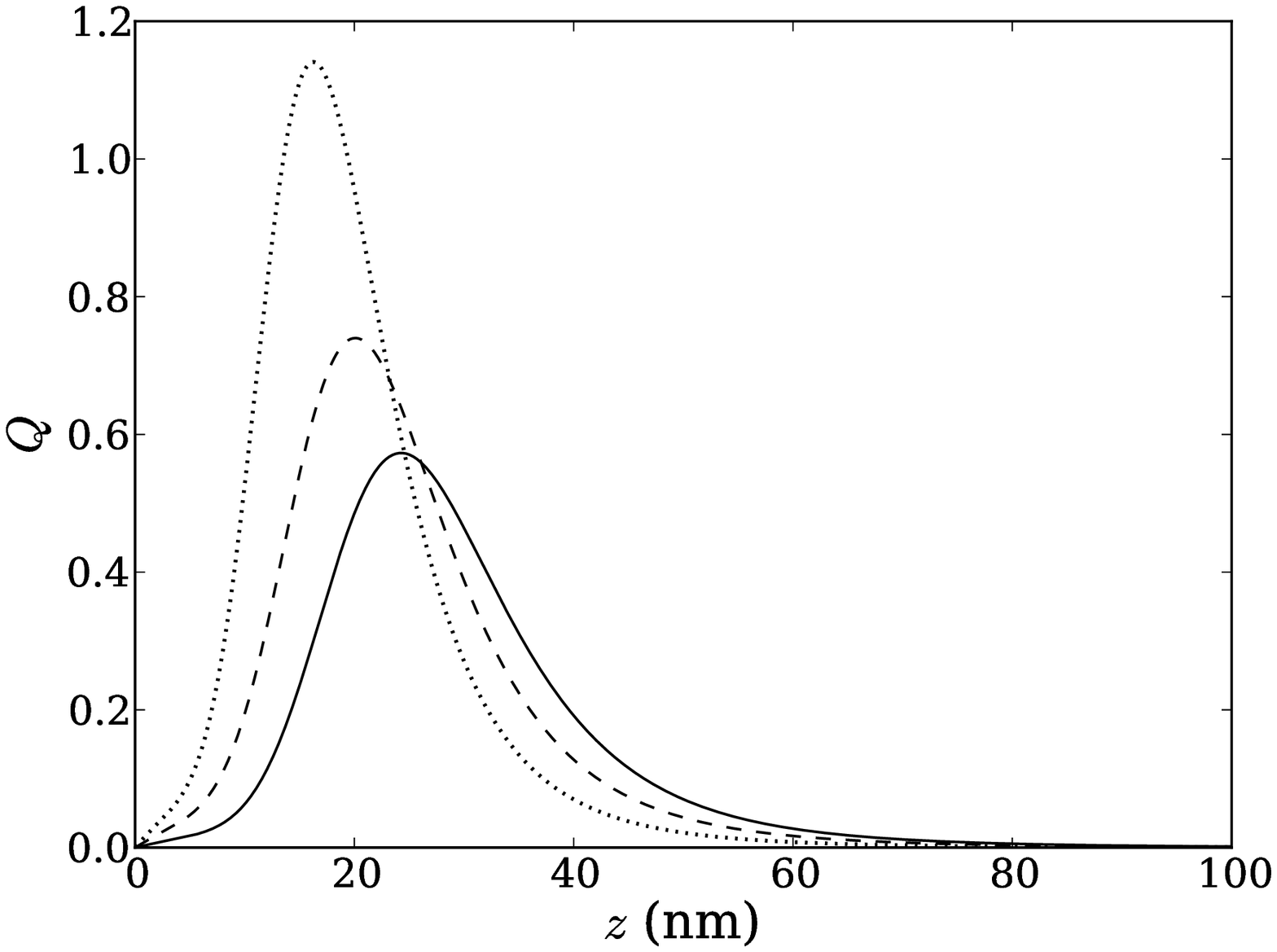}\hspace{-.5cm} \includegraphics[width=6.5cm]{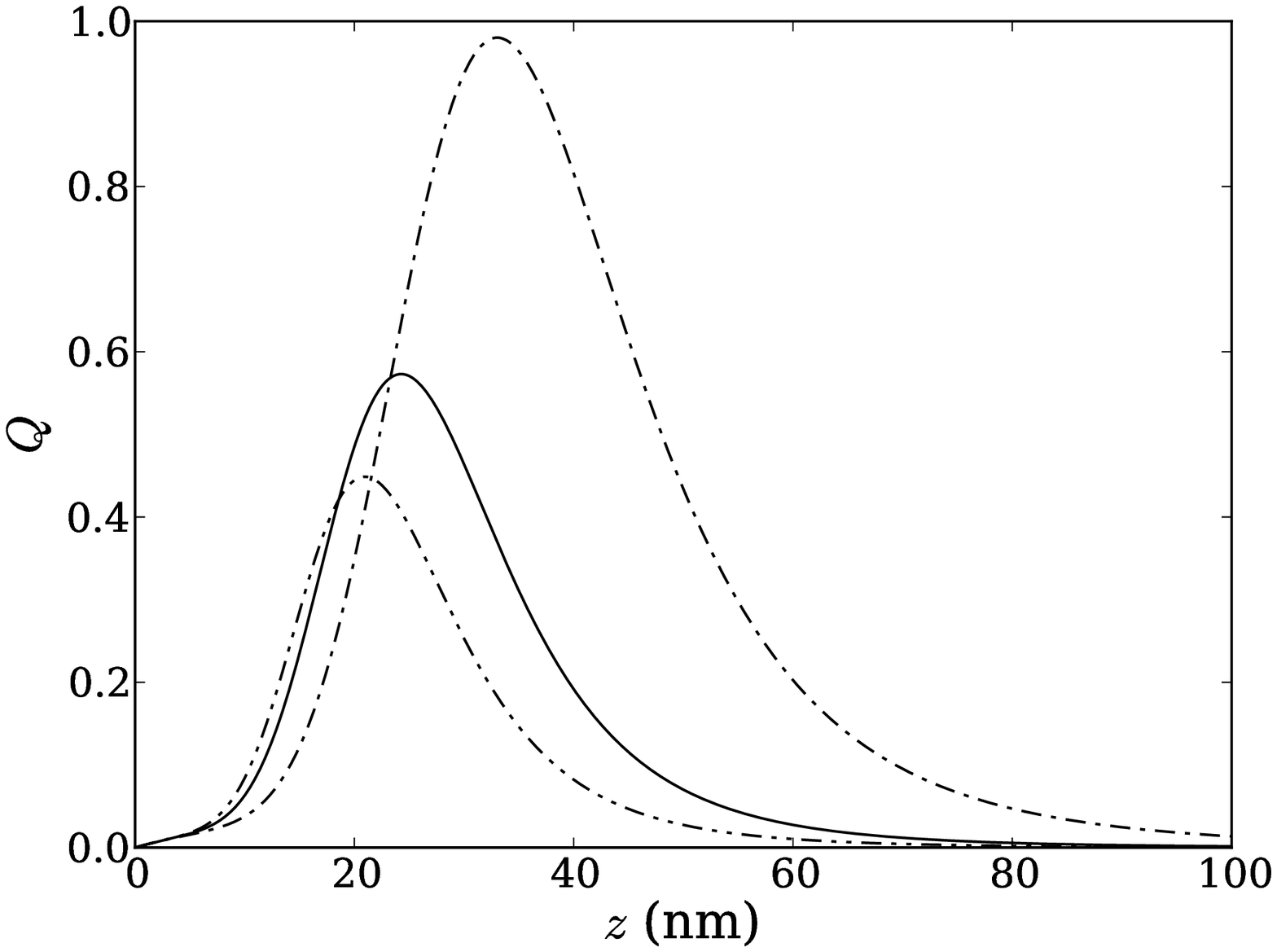}
\caption{\label{badlands} Left: Badlands function $Q(z)$ as a function of distance to the
surface for $\Hb$ dropped from $h=10$~cm on bulk mirrors; from bottom right to top left, perfect conductor (full line),
silicon (dashed line), silica (dotted line).\newline
Right: Badlands function $Q(z)$ as a function of distance above a perfectly conducting mirror for $\Hb$ dropped from $h=10$~cm (dash-dot line), $h=30$~cm (full line) and $h=50$~cm (dash-dot-dot line).
}
\end{figure}

The plots in figure \ref{badlands} clarify the subtle dependence of the magnitude and position of quantum reflection on the potential and on the energy.
The left plot shows that if the energy is fixed, reflection on weaker potentials happens closer to the surface, where the potential is steeper, so that the deviation from WKB is greater and greater. 
On the right hand side, the potential is fixed and the energy varied between the three curves. As expected the non-classical behavior is inhibited as the energy increases, even though the reflection region is pushed towards the surface. Indeed, the effect of increasing the energy outweighs the fact that reflection occurs on a steeper potential.

\section{Increasing quantum reflection}

In the context of GBAR quantum reflection is first a bias to be mastered. More generally however, efficient quantum reflection opens exciting possibilities for trapping and guiding antimatter with material walls. 
For example, antihydrogen held in a gravity field by quantum reflection settles in gravitational quantum states. If the lifetime of such states is long enough, their study could lead to orders of magnitude improvements on the determination of $\gb$\cite{voronin2011}. 
Quantum reflection could also be used in GBAR to reduce the initial velocity distribution of atoms which limits the precision of the experiment. Atoms would pass between two disks, a bottom disk with a smooth surface to reflect slow enough $\Hb$ atoms and a rough top disk that scatters atoms non-specularly, leading to their loss. Atoms that are fast enough to rise in the gravity field and touch the top disk are therefore eliminated\cite{shaper}.

In the light of the previous section's discussion, efficient quantum reflection is obtained for slow particles and weak CP interactions\cite{judd2011}. 
To weaken the CP interaction, one strategy is to remove matter from the mirror, so as to reduce its coupling to the field. This can be done by going to thinner or to less dense mirrors. In the present section we study reflection on thin slabs, graphene layers and nanoporous materials. 
The latter incorporate a significant fraction of gas or vacuum and their characteristic pore/grain size is in the 1-10 nm range. Here we give results for silica aerogels, powders of diamond nanoparticles and porous silicon\cite{dufour2013-a}.

These various materials can all be treated within the scattering theory of Casimir forces by including the suitable optical response in the calculations:
\begin{enumerate}
  \item slabs of finite thickness $d$ have smaller field reflection amplitudes than the corresponding bulks, in particular they are transparent to large wavelengths. The CP potential is unaffected at distances smaller than $d$ but it falls off more rapidly at large distances: $V(z) \underset{z\gg \lambda,d}{\to}-C_5/z^5$;
  \item field reflection amplitudes on a graphene sheet are given by the Dirac model\cite{dufour2013};
  \item nanoporous materials are described by the Bruggeman effective medium theory\cite{bruggeman1935}, we can expect approximately correct results for processes involving scales larger than the scale of inhomogeneities in the material\cite{dufour2013-a}.
\end{enumerate}

\begin{table}[h]
\caption{Quantum reflection probability and lifetime of the first gravitational quantum states of $\Hb$ above various material surfaces.}
{\label{ref-life}\begin{tabular}{C{5cm} |C{3cm} | C{3cm}} \hline \hline
material & reflection probability ($E=mg\times30$~cm) & lifetime of first gravitational quantum states (s)\\\hline
perfect conductor         & 5\%   & 0.11 \\ \hline
bulk silicon              & 9\%   & 0.14 \\ \hline
bulk silica               & 18\%  & 0.22 \\ \hline
5 nm silica slab          & 27\%  & 0.33 \\ \hline
graphene                  & 44\%   & 0.55 \\ \hline
nanodiamond powder (porosity 95\%) &   &   0.89   \\ \hline
porous silicon (porosity 95\%)     &   &   0.94   \\ \hline
silica aerogel (porosity 98\%)     &   &  4.6 \\
 \hline\hline
\end{tabular}
}
\end{table}

In table \ref{ref-life} we give reflection probabilities for $\Hb$ with energy $E=mg\times30$~cm on various materials. In the case of nanoporous materials, the use of an effective medium approximation is valid only if the atoms are reflected at a distance larger than the medium's inhomogeneities. This is not the case for atoms with energy $E=mg\times30$~cm, who would come within a few nanometers of the surface. Instead we give the lifetime of the first gravitational quantum states. These are metastable state in the potential ``well'' formed by gravity and quantum reflection. Their lifetime is limited by the probability of annihilation on the surface and can be shown to be the same for the lowest energy states\cite{voronin2011}. This lifetime increases dramatically for porous materials compared with bulk materials, reaching several seconds for silica aerogel.

\section{Conclusion}

The scattering approach to Casimir forces gives realistic estimates of the CP potential between $\Hb$ and a variety of material mirrors. Calculations of quantum reflection probabilities on such potentials have shown that substantial reflection was to be expected in the GBAR experiment. Also, we explained why weaker potentials lead to higher reflection and showed how this could be used to manipulate and study antimatter.

\bibliographystyle{unsrt}
\bibliography{biblio}

\end{document}